\documentclass[letterpaper,12pt]{article}
\usepackage{fullpage}
\usepackage{amssymb}
\usepackage{amsmath}
\usepackage{graphicx}
\usepackage{subfigure}

\title{Customer Appeasement Scheduling\footnote{Technical Report TR-10-18, School of Computer Science, Carleton University. This work is partially supported by NSERC.}}
\author{Mohammad R Nikseresht \and Anil Somayaji \and  Anil Maheshwari}

\begin{document}

\date{}
\maketitle

\begin{abstract}
Almost all of the current process scheduling algorithms which are used in modern operating systems (OS) have their roots in the classical scheduling paradigms which were developed during the 1970's.
But modern computers have different types of software loads and user demands.
We think it is important to run what the user wants at the current moment.
A user can be a human, sitting in front of a desktop machine, or it can be another machine sending a request to a server through a network connection.
We think that OS should become intelligent to distinguish between different processes and allocate resources, including CPU, to those processes which need them most.
In this work, as a first step to make the OS aware of the current state of the system, we consider process dependencies and interprocess communications.
We are developing a model, which considers the need to satisfy interactive users and other possible remote users or \emph{customers}, by making scheduling decisions based on process dependencies and interprocess communications.
Our simple proof of concept implementation and experiments show the effectiveness of this approach in the real world applications.
Our implementation does not require any change in the software applications nor any special kind of configuration in the system, Moreover, it does not require any additional information about CPU needs of applications nor other resource requirements.
Our experiments show significant performance improvement for real world applications. For example, almost constant average response time for \emph{Mysql} data base server and constant frame rate for \emph{mplayer} under different simulated load values.

\end{abstract}

\section{Introduction}
\subsection{Motivation}
\label{sec-introduction-motiv}
Almost all of the current process scheduling algorithms, which are used in modern operating systems (OS), have their roots in the classical scheduling paradigms which were developed during the 1970's.
But today's computers have different types of software loads and user demands.
It is not important to maximize CPU utilization, as most modern machines, either desktops or servers, have multiple cores/CPUs and most of the time they have idle CPU cycles.
It is not important to minimize job turn around time, because most machines are not running CPU intensive jobs. 
Most machines have a varying load pattern which depends on the requests coming from local and/or remote users.
Often it may not be important to maximize system throughput.
Typically throughput is defined in terms of non-interactive jobs submitted to a machine, while most modern tasks have some form of interaction with users.
What is important is to run what the user wants at the current moment.
A user can be a human, sitting in front of a desktop machine, or it can be another machine sending a request to a server through a network connection.
As we are looking at a wider range of users, we call them \emph{customers}.
In our view, local or remote customers may send different requests to a system.
Due to rapid changes in customer demands and requests, the need for CPU time among different processes or groups of processes changes rapidly.
Some OSs and authors have framed part of this problem as  the user interactivity problem and have addressed it in several ways, but none of them have presented an easy to use solution \cite{PrOutPutProduction,freeBSD,Linux2.6Interactivity,windowsInternal,holistic}.
These solutions usually need some form of configuration and/or input from the user, or need additional information from applications.
We will elaborate on this in Section \ref{sec-realted-work}.

We believe a key change in OS resource management is to make them aware of what the applications are doing on top of them.
In other words, we think OS should become intelligent to distinguish different processes and allocate resources, including CPU, to those processes which need them most.
We think the processes that need more resources are the ones which are \emph{externally observable} at the time of scheduling.
If a customer is waiting for a response from a process, then we say that, this process is \emph{externally observable}.
If this process is waiting for a service from another process, then the second process is also \emph{externally observable}. 
We believe that as a first step to make OS aware of what is happening in the system,  process dependencies and interprocess communications should be considered.
Unfortunately, commodity OSs do not support process dependency detection or interprocess communication detection.
Although, OS kernel usually has some information about interprocess communications and process dependencies, they are generally dispersed in various unrelated kernel data structures, and the kernel does not use those information to make any process scheduling decisions or any other resource allocation decisions.

In this study we are developing a model which considers the need to satisfy interactive users and other possible remote users or customers. This model makes scheduling decisions based on process dependencies and interprocess communications.
We want to develop a scheduling algorithm which tries to minimize a user's dissatisfaction or unhappiness.
We call this \emph{customer appeasement} as it is not possible to make every customer satisfied specially under heavy loads by running all processes fairly.
A scheduling policy resulting from customer appeasement model is not a fair scheduling policy as it tries to find more important processes and give them more priority.
This goal is achieved by a model which tracks process dependencies and communications using scalar values assigned to processes, customers and the whole system.
Our simple proof of concept implementation and experiments show the effectiveness of this approach in real world applications.
Our implemention does not require any specific change in the software applications or in the configuration of the system. Moreover it does not require any additional information about CPU needs of applications and other resource requirements.

\subsection{Related Work}
\label{sec-realted-work}
As mentioned in Section \ref{sec-introduction-motiv}, as far as we know, there are many studies which try to solve interactive or multimedia applications scheduling problems, but none of them has a broader view of finding the optimal scheduling solution based on a well defined criteria for all applications, specially under heavy load.

Most commodity OSs use some heuristics based on process execution/sleeping behavior to detect interactive processes to increase their priority and reduce their latency.
Windows \cite{windowsInternal} and FreeBSD \cite{freeBSD} use multi-level feedback queue schedulers.
In this scheme CPU-bound processes receive lower priorities and processes blocked waiting for I/O receive higher priorities.
The Linux Vanilla or O(1) scheduler \cite{Linux2.6Interactivity} (used in kernels before 2.6.23) has a similar mechanism.
Processes with longer sleep times and shorter execution times are identified as interactive and receive higher priorities.
Windows \cite{windowsInternal} adds more intelligence by differentiating processes waiting on different devices.
For example, processes waiting on keyboard receive higher priority than those waiting on a disk.
Etsion et al. and Yan et al. \cite{PrOutPutProduction,holistic} show that depending only on execution behavior is not adequate to distinguish interactive processes properly.
Ingo Molnar, the designer of Linux CFS scheduler \cite{CFS2-IMolnar,CFS-Mingo} tries to mitigate this problem by not depending too much on process execution/sleeping behavior.
CFS scheduler doesn't change interactive processes' priority any more,
it only inserts them in front of the run queue every time an interactive process wakes up \cite{CFS2-IMolnar,CFS-Mingo} (also see Section \ref{CFS-unhap-calc}).

Windows \cite{windowsInternal} also uses ``windows system input focus'' as a measure of user interaction and it increases the priority of a process which has the input focus. 
Using input focus may help to improve interactivity performance but has several problems. If a user is running multiple interactive programs, for example an audio player and a web browser, while he/she is browsing the web and input focus is on the web browser, the user still wants the audio player to play the music well.
Input focus mechanism also might not be usefull if a user interacts with the system through the network.

Etsion et al. \cite{PrOutPutProduction} use process display output production as a means of detecting interactive and multimedia applications.
They schedule processes based on their display output production in a way that all processes have a chance to produce display output at the same rate.
That might be usefull for multimedia applications where, for example, all video applications play at the same frame rate regardless of their window size.
This approach only addresses desktop applications as any network user has no display access.
Also, it might be possible that a compute intensive job creates a huge amount of disply output and receives an increase in its priority while it actually is not an interactive application.

Some researchers and OSs, allow real time or interactive processes to specify their CPU requirements and time constraints.
For example in Mac OS X \cite{OSX2006}, a real time process may ask for a specific CPU requirement.
Yang et al. (RedLine) \cite{YangRedline08} use almost the same principles and treat interactive processes like real time processes. In RedLine processes can ask for a specific CPU and other resource requirements. RedLine also has an admission mechanism which may not allow the process to execute as an interactive process if the system does not have enough resources as requested by the process.

Zheng et al. have an implementation called SWAP \cite{swap} which recognizes process dependencies but it does not distinguish interactive or any other type of processes which might need increased priority.
It only tracks process dependencies based on system calls and prevent a high priority process being blocked by a low priority process that has locked a resource needed by the high priority process (priority inversion problem). 

Zheng et al. work called RSIO \cite{RSIO_ZhengN10} has the most similarities with our work.
RSIO looks at process I/O patterns as a way of detecting interactive processes.
It also tries to identify other processes involved in a user activity and provide a scheduling policy to improve interactive performance. This policy is based on access patterns to I/O devices. RSIO needs a configuration file that defines which I/O devices should be monitored to detect interactive processes. It has also a relatively complicated heuristic mechanism to detect processes involved in a user interaction. 

\subsection{Contributions}
\label{sec-intro-contrib}
The work presented in this paper has major differences from all of the previous work.
\begin{enumerate}
 \item We develop a model (customer appeasement model) with a criteria which tracks process dependencies and customer requests.
This model gives us the ability to compare different schedulers analytically, and develop new scheduling policies based on the analytical results.

\item Our model considers all of the process communications and dependencies in the system which are indications of a customer request. Other systems e.g. RSIO \cite{RSIO_ZhengN10} typically consider a subset of communications related to a subset of I/O devices.



\item The customer appeasement model objective is to improve the performance of any externally observable processes specially under heavy background loads, this includes traditional interactive processes, i.e. desktop and multimedia applications.



\item We have a simple proof of concept implementation which does not need any configuration file, or process specification information.
It does not require any changes in the software applications either.
It automatically and without any user assistance detects those processes which need more resources as defined in this paper, and increases their priority.
This is in contrast to other work such as Redline \cite{YangRedline08}, RSIO \cite{RSIO_ZhengN10} or as allowed by OS X \cite{OSX2006}, which require some form of configuration or process resource specifications from user.

\item Our experimental results show significant performance improvements for both interactive applications and server processes such as \emph{Apache} web server \cite{apache} and \emph{Mysql} \cite{Mysql} data base servers. For example, we observed almost constant average Mysql response times, and almost constant frame rate for \emph{mplayer} under different (simulated) background loads. 

\item One of our goals is to make the implementation simple, easily portable to different Linux kernels and distributions, and easy to use for a novice user.
In order to achieve this, we use SystemTap \cite{SystemTap}.  SystemTap is a diagnostic tool, but it has made our implementation a simple script which can be run on any SystemTap equipped distribution with a compatible kernel without the need to recompile and install a new kernel.
Our script has been tested on kernel versions 2.6.31, 2.6.32, 2.6.35, but should be compatible with any kernel that has a recent version of CFS scheduler.
\end{enumerate}

The rest of this paper is organized as follows:
We describe the customer appeasement  model and its basic definitions in Section \ref{sec-unhappyModel}. 
In Section \ref{sec-comparison-theory}, we compute the unhappiness values for two simple scenarios for some of the classical and modern OS schedulers.
In Section \ref{sec-unhappy-shced-propos}, we propose an algorithm to use unhappiness values to change process scheduling.
In Section \ref{sec-requestbased}, we explain our simplified request based priority elevation technique.
We give a more detailed explanation on the implementation in Section \ref{sec-implemetation}.
In Section \ref{sec-experiments}, we present our experimental results, concluding remarks are presented in Section \ref{sec-conclusion}.

\section{The Customer Appeasement Model}
\label{sec-unhappyModel}
In this section we introduce the customer appeasement model and explain the parameters and variables in detail.

\subsection{Definitions}
In the customer appeasement model we use the following terms, notations and definitions:

\begin{description}
 \item[Process:] A process $P$ is any software entity inside the system which can be scheduled to run on a CPU by the OS. (This definition includes tasks, threads or light weight processes.)

 \item[Customer:] A customer $C$ is any outside entity which can send requests to  processes in the system.
Customers are independent from each other.
We may distinguish between local and remote customers.

 \item[Direct Request:] A request $R$ is any type of input from a customer to a process. $R^{k}_{i\rightarrow j}$ denotes the $k^{th}$ request from customer $C_{i}$ to process $P_{j}$. 
 
\item[Indirect Request:] A process may receive a request from a customer indirectly. This happens when a process which has received a direct request from a customer in turn, sends a service request to another process.

 \item[Weight of a request:] $w^{k}_{i}$ is the weight or importance of the request $R^{k}_{i\rightarrow j}$ for the customer $C_{i}$.
This may be measured or inferred from customer's behavior.

 \item[Customer weight:] $W(c_{i})$ is the parameter which is used to distinguish between different customers. It represents weight or importance of customer $c_{i}$ for the system.

\item[Unhappiness:] $u^{k}_{i \rightarrow j}$ is an integer value related to the time delay that customer $C_{i}$ experiences as a result of sending the request  $R^{k}_{i\rightarrow j}$ to the process $P_{j}$. 
\end{description}

\subsection{Computation of Unhappiness Value}

The amount of unhappiness assigned to a process due to a request, changes according to the rules explained in this section. 
The request might have been sent either directly or inderectly through another process to $P_{j}$.
In the simplest situation, the unhappiness for process $P_{j}$ at any moment is defined as the elapsed time since the moment process $P_{j}$ receives $R^{k}_{i\rightarrow j}$ minus the amount of CPU time that $P_{j}$ has been allocated.
When $P_{j}$ sends a response back to the customer $C_{i}$, then $u^{k}_{i \rightarrow j}$ is set to zero. Observe that:

\begin{enumerate}
 \item The unhappiness value $u^{k}_{i\rightarrow j}$ increases as time passes.

 \item $u^{k}_{i\rightarrow j}$ is decreased by the amount of time that process $P_{j}$ runs on a CPU processing request $R^{k}_{i\rightarrow j}$.

 \item When process $P_{j}$ requests a service from another process $P_{s}$ and blocks, then $u^{k}_{i \rightarrow j}$ is divided between $P_{j}$ and $P_{s}$ as follows:

\begin{align}
\label{eq-unhapValAlpha} 
&u^{*k}_{i \rightarrow j} = \alpha u^{k}_{i \rightarrow j} \nonumber \\
&u^{k}_{i \rightarrow s} = (1-\alpha) u^{k}_{i \rightarrow j} 
\end{align}
where $0 \leq \alpha < 0.5$  is a system parameter, and it determines the amount of unhappiness passed to a service process when an indirect request is sent to such a process.
Its exact value should be determined based on experiments during a specific implementation.

 \item The new unhappiness value $u^{*k}_{i \rightarrow j}$ for $P_{j}$ does not change while process $P_{j}$ is blocked waiting for a service from other processes. But $u^{k}_{i \rightarrow s}$ will increase by time and in general follows the same rules for unhappiness computation.

 \item When the service process $P_{s}$ finishes its processing and returns a response to $P_{j}$, effectively unblocking $P_{j}$ by giving the requested service to it, the value of $u^{k}_{i \rightarrow s}$ is passed back to $P_{j}$ and is added to its previous unhappiness value.
 We call this new unhappiness value $u^{**k}_{i \rightarrow j}$.
 At this time the unhappiness value assigned to service process is reset to zero:

\begin{align}
&u^{**k}_{i \rightarrow j} = u^{*k}_{i \rightarrow j} + u^{k}_{i \rightarrow s} \nonumber \\
&u^{k}_{i \rightarrow s} = 0
\end{align}

\end{enumerate}


We can compute the total unhappiness for a request, a customer ($U^{c_{i}}$) or the whole system ($U$).
The total unhappiness for a request $R^{k}_{i \rightarrow j}$ is computed as the following summation which indicates the total unhappiness that customer $C_{i}$ experiences as a result of sending request $R^{k}_{i \rightarrow j}$ to process $P_{j}$ and all other delays that are caused as process $P_{j}$ waits for services from 
other processes. 

\begin{equation}
\label{eq-totalUnhap-Req}
U^{R_{i}^{k}}= W(c_{i}) w^{k}_{i} \sum_{j=0}^{N-1}u^{k}_{i \rightarrow j}
\end{equation}

The total unhappiness for customer $C_{i}$ is computed using Equation \ref{eq-totalUnhap-cus}. In this equation $R$ is the total number of requests sent by $C_{i}$ and $N$ is the total number of processes in the system:

\begin{equation}
\label{eq-totalUnhap-cus}
U^{c_{i}}= W(c_{i}) \sum_{j=0}^{N-1} \sum_{k=0}^{R-1}w^{k}_{i}u^{k}_{i \rightarrow j}
\end{equation}

The system unhappiness due to all requests from all customers is computed using Equation \ref{eq-totalUnhap-sys}:

\begin{equation}
\label{eq-totalUnhap-sys}
 U=\sum_{i=0}^{M-1}W(c_{i}) \sum_{j=0}^{N-1} \sum_{k=0}^{R-1}w^{k}_{i}u^{k}_{i \rightarrow j}
\end{equation}

The objective of the scheduling algorithm should be to minimize the system's unhappiness $U$ at any time.

\section{Unhappiness Values in different Scheduling Algorithms}
\label{sec-comparison-theory}
In order to find out  how some of the scheduling algorithms perform under the customer appeasement model, we compute the unhappiness value that they cause for a request sent by a customer to a system in two specific scenarios.
We compute the unhappiness value for the following schedulers, and simplify final values as much as possible so that the results are comparable.

\subsection{Round-Robin Scheduling}
The Round-Robin (RR) Scheduler is a simple preemptive scheduling algorithm which was used in time sharing systems \cite{OSTanenbaum}.
It is still used as part of some modern scheduling algorithms, for example it is part of Linux real time (RT) scheduling class. 
Round-Robin scheduler gives each process a time slice or time quantum $q$, if a process releases the CPU before $q$ is finished,
then the scheduler runs the next process in the ready queue.
If a process needs more time and finishes its time quantum, then the scheduler preempts the process, inserts the process at the tail
of the ready queue, and schedules the next process from the head of the ready queue.
This new running process also receives a time quantum $q$.

Now we consider a simple case and compute the minimum and a typical unhappiness values caused by a request in a system with RR scheduler.
The minimum and typical unhappiness values might happen in the best case and a typical case scenarios respectively.
We assume that there are $N-1$ running processes in the run queue.
We assume that there is a process $P_j$ in the sleeping state waiting to receive a request. There is only one customer, and the system parameter $\alpha$ is set to zero ($\alpha$ = 0).
This customer sends a request to the sleeping process $P_{j}$ and waits for the response.
Now the OS wakes up process $P_{j}$ and inserts it to the end of the ready queue.
Assume that the $N-1$ running processes stay running all the time, and they use all of their time quanta , so $P_{j}$ waits in the
ready queue for $q(N-1)$ seconds before it runs on the CPU. If it can finish processing and return a response to the customer during its first time slice $q$, then $q(N-1)$ is the amount of unhappiness during this transaction.
If it can't return a response to the customer during this time and needs a total of $Z$ time quanta to finalize this transaction and 
return a response to the customer, then the maximum unhappiness will be:

\begin{equation}
\label{eq-unhap-RR}
 U^{R}=Zq(N-1) -(Z-1)q = q(Z(N-2)+1)
\end{equation}

In practice it is possible that process $P_{j}$ blocks and waits for services from 
other processes.
Assume that it waits for a service from process $P_{s}$ after $Z_{1}$ time quanta, and $P_{s}$ needs $Z_{2}$ time quanta to process $P_{j}$'s request and return a response.
$P_{j}$ may also need another $Z_{3}$ time quanta to return a response to the customer.
We assume that $P_{s}$ is in the sleeping state prior to receiving a request from $P_{j}$, that means, there are always $N$ running processes, because when $P_{j}$ blocks and sleeps, $P_{s}$ wakes up and is in the running state.
Then the amount of unhappiness experienced by the customer due to the request $R$ will be:
\begin{align}
U^{R} &=q(Z_{1}(N-2)+1)+q(Z_{2}(N-2)+1)+q(Z_{3}(N-2)+1) \nonumber \\
&= q ((Z_{1}+Z_{2}+Z_{3})(N-2)+3)
\end{align}

Here the unhappiness value consists of three terms. 
The first term is the aggregated unhappiness caused by delays in the execution of $P_{j}$ at the time it blocks waiting on $P_{s}$.
The second term is the amount of unhappiness caused by delays during the execution of $P_{s}$, and the last term of the unhappiness value reflects the delays of running $P_{j}$ after it receives the response from $P_{s}$ until it sends the final response to the customer.
So the best case and a typical case scenarios with a Round-Robin scheduler results in the following minimum and typical unhappiness values:
\begin{align}
\label{eq-RondRobin-final}
 &U^{R}_{min} = q(Z(N-2)+1)\\
 &U^{R} = q((Z_{1}+Z_{2}+Z_{3})(N-2)+3)
\end{align}

Note that the unhappiness value related to running the service process $P_{s}$ is set to zero once it sends the response back to $P_{j}$.

\subsection{Multilevel Feedback Queues}
In this subsection we perform the same analysis for a basic multilevel feedback queue scheduling. 
Many UNIX OSs such as FreeBSD \cite{freeBSD} utilize some form of a multilevel feedback queue scheduler. Windows also has a multilevel feed back queue scheduler \cite{windowsInternal}.
As another example the $O(1)$ or Vanilla scheduler in Linux kernels before 2.6.23 is in fact a multilevel feedback queue with many heuristics involved in moving tasks between diffrent queues and detecting interactive processes \cite{Linux2.6Interactivity}.

Assume a basic multilevel feedback queue scheduling algorithm with $m$ queues called $Q_{0}$ to $Q_{m-1}$. The processes in each queue can run on the CPU for a multiple of time quantum $q$. The amount of time for $Q_{i}$ is computed as $2^{i}q$. Each process is first placed in $Q_{0}$. After it receives its CPU share in $Q_{0}$, if it needs more time it is then placed in the next queue $Q_{1}$ and so on. Each queue has an absolute priority relative to the next queue, meaning that processes in $Q_{i+1}$ do not execute until all processes in $Q_{i}$ receive their CPU share and $Q_{i}$ becomes empty. So in this algorithm, processes which need more CPU time, lose their priority as time passes but receive a,larger time quantum when they run on the CPU.

Now assume an interactive process wakes up and receives a request from a customer.
Also assume that there are a total of $a_{i}$ processes in $Q_{i}$. 
Assume that the interactive process needs $Zq$ processing time to finish processing and return a response to the customer, and no other processes will enter the running queues during this time.
This means that the interactive process is inserted to the end of $Q_{0}$, receives its CPU time after waiting for other processes in this queue, then it is pushed to the next queue and so on.
Assume $Q_{x}$ is where it receives the final amount of CPU time that it needs to finish  processing the request and return a response to the customer.
The amount of unhappiness that the customer experiences is computed as:

\begin{align}
\label{eq-mulfeedQ}
 U^{R} &= (a_{0}-1)q + 2(a_{1}-1)q + ... + 2^{x}(a_{x}-1)q-(q+2q+...+2^{x-1}q) \nonumber \\
&= q(\sum_{i=0}^{x}2^{i}(a_{i}-1)-\sum_{i=0}^{x-1}2^{i})
\end{align}
In this equation the first summation indicates the amount of delays that the interactive process encounters waiting in ready queues, and the second summation indicates the amount of CPU time it has received. 
We can compute $x$ by solving this equation $\displaystyle\sum_{i=0}^{x}2^i=Z$ and then simplifying the minimum unhappiness value as follows:

\begin{align}
\label{eq-mlfq-x-and-urmin}
&x=\log_{2}(Z+1) -1\\
&U^{R}_{min}=q(\sum_{i=0}^{\log_{2}(Z+1)-1}2^{i}(a_{i}-1)-2^{\log_{2}(Z+1)-1})
\end{align}

We can also examine a more complicated scenario as in the previous subsection.
Assume that $P_{j}$ wakes up and receives a request. 
It then spends $Z_{1}$ second to partially process this request and send a request to a service process $P_{s}$.
Now $P_{s}$ needs $Z_{2}$ seconds to provide the service to $P_{j}$.
After $P_{j}$ receives the service, it needs another $Z_{3}$ seconds to finalize its processing and return a request to the customer.
Please note that in this scheduler each time a process wakes up, it is inserted to the end of $Q_{0}$.
We compute unhappiness values assuming that the system parameter $\alpha$ is set to zero ($\alpha = 0$) see Section \ref{sec-unhappyModel}:

\begin{align}
\label{eq-multyFQ-final}
 U^{R} &=q(\sum_{i=0}^{\log_{2}(Z_{1}+1)-1}2^{i}(a_{i}-1)-2^{\log_{2}(Z_{1}+1)-1})\nonumber
 + q(\sum_{i=0}^{\log_{2}(Z_{2}+1)-1}2^{i}(a_{i}-1)-2^{\log_{2}(Z_{2}+1)-1})\\
 &+ q(\sum_{i=0}^{\log_{2}(Z_{3}+1)-1}2^{i}(a_{i}-1)-2^{\log_{2}(Z_{3}+1)-1})
\end{align}

Please note that the total unhappiness value consists of three terms. The first term is the result of delays during the first part of processing the request by $P_{j}$. The second term is caused when the service process $P_{s}$ is in the run queue, and the last term is associated with the waiting when $P_{j}$ prepares the final response for the customer.
This scenario can be a typical situation while it is possible to have even more complex cases where $P_{j}$ may need more services from $P_{s}$ or other service processes. This leads to higher unhappiness values.

\subsection{Linux CFS Scheduler}
\label{CFS-unhap-calc}
The Completely Fair Scheduler or CFS for short \cite{CFS2-IMolnar,CFS-Mingo} was introduced in Linux kernel version 2.6.23.
As its developer explains \cite{CFS2-IMolnar,CFS-Mingo}, ``it is designed to basically model an ideal, precise multi-tasking CPU on real hardware''.

It allways tries to share the CPU fairly between the current processes in the run queue.
In other words if there are $N$ processes in the run queue, it promises to run each process with $\frac{1}{N}$ of the CPU power.
In order to achieve this goal, CFS scheduler keeps track of a variable called virtual run time ($vruntime$) for each task.
It is a weighted run time for each task.
CFS uses an R-B tree to choose the next task to run on the CPU.
It simply chooses the left most task in the tree which has the lowest $vruntime$ value.
If a new process enters the run queue, CFS manipulates its $vruntime$ value such that the new arriving process goes to the right of the R-B tree.
This is to make sure that it can keep its promised wait time to the current running processes.
CFS also gives an advantage to the sleeping processes.
If a process sleeps less than a threshold time interval then CFS changes its $vruntime$ value such that it goes to the left most position in the R-B tree when it wakes up.
Assuming that an interactive task is a short sleeper, this will lead to better response times for interactive tasks.
CFS does not have a fixed time slice.
At the time it runs the next task it gives the task a time slice which is computed as follows:
\begin{equation}
\label{eq-CFS-timeSlice}
 Time slice = q(N) =\frac{sch\_lat}{N}
\end{equation}
In this equation $sch\_lat$ is a CFS constant value and $N$ is the number of tasks in the run queue.
CFS stretches this time slice if the number of running tasks increases beyond a system threshold. 
$q(N)$ is the basic time slice in CFS if process weights and nice values are ignored.

Another intresting property of CFS scheduler is the way nice values work.
Nice values change the weight of a task, which means that they change the $vruntime$ of a task.
If task weights and nice values are considered then CFS changes the CPU share of each task based on the following relations:
\begin{align}
\label{eq-taskCPUSahre-basedOnNice}
&q(P_{j},N) = \frac{w_{j}sch\_lat}{\sum_{i=1}^{N}w_{i}} \\
&wnice_{0}=1024 \nonumber \\
&wnice_{i-1} = 1.25wnice_{i} \nonumber 
\end{align}
$w_{i}$ in equation \ref{eq-taskCPUSahre-basedOnNice} is the weight of $P_{j}$ assigned by CFS, and changes directly based on the $P_{j}$'s nice value. 
For example if there are two tasks $A$ and $B$ running on a single CPU machine and both have a default nice value of $0$ then each receives 50\% of the total CPU time.
If task $A$'s nice value is changed to $-1$ then it receives 55\% of the CPU time and task $B$ receives 45\% of the CPU power.

Now we compute the unhappiness values experienced by a customer in a system with CFS scheduler.
Assuming that all running processes have their default nice value of zero, we have:
\begin{align}
\label{eq-QnNiceZero}
q(P_{j},N) = \frac{1024\hspace{10pt}sch\_lat}{{1024\hspace*{10pt}N}}= \frac{sch\_lat}{N}=q(n)
\end{align}

Assume that there are $N-1$ running processes in the run queue and an interactive task $P_{j}$ is sleeping.
Assume a customer sends request $R$ to $P_{j}$.
$P_{j}$ wakes up and enters the run queue.
Now we compute the minimum unhappiness that may be experienced by the customer.
In the best case scenario, it is possible that CFS changes the $vruntime$ of $P_{j}$ such that it is placed in the left most leaf of the R-B tree and it then may preempt the current running process.
Please note that now the number of running processes is $N$.
Assume that $P_{j}$ needs $\tau$ seconds to finish processing the request.
If $\tau \leq q(N)$
then $P_{j}$ can finish processing the request without interruption and returns a response to the customer and the total unhappiness value is zero.
\begin{equation}
\label{eq-unhap-CFS-final-min}
 U^{R}_{min} = 0
\end{equation}

Now we assess a typical scenario where $\tau > q$ 
and  the  interactive task blocks to receive a service from service task $P_{s}$.
We assume that the system parameter $\alpha = 0$, and when $P_{j}$ blocks, it transfers all of its unhappiness to $P_{s}$.
At this time $P_{s}$ enters the run queue so the total number of running jobs does not change.
Additionally we also assume that no other task enters or leaves the run queue while the request $R$ is being serviced.
So the total number of running tasks is always $N$.
Assume that $P_{j}$ needs $\tau_{1}$ seconds for processing the request $R$  before blocking and requesting service from $P_{s}$, and $P_{s}$ needs $\tau_{2}$ seconds to return the requested service to $P_{j}$, and $P_{j}$ needs another $\tau_{3}$ seconds to return a response to the customer.
To simplify the computation we use 
$\tau = \tau_{1} + \tau_{2} + \tau_{3}$. 
The total unhappiness experienced by customer due to this request is:
\begin{align}
\label{eq-CFS-final-typical}
U^{R} &= (\frac{\tau_{1}}{q(N)}-1)(N-1) - \tau_{1} + (\frac{\tau_{2}}{q(N)}-1)(N-1) - \tau_{2} + (\frac{\tau_{3}}{q(N)}-1)(N-1) - \tau_{3} \nonumber \\
&= (N-1)(\frac{1}{q(N)}(\tau_{1}+\tau_{2}+\tau_{3})-3)-(\tau_{1}+\tau_{2}+\tau_{3})\nonumber \\
&= (N-1)(\frac{1}{q(N)}\tau -3) -\tau
\end{align}


\section{Scheduling Based on Unhappiness Values}
\label{sec-unhappy-shced-propos}
Up to this point, we have developed a model which can give us an indication of how an OS performs regarding the requests it receives from different customers.
Interstingly, the way we have defined unhappiness, and compute its value, and the way it is inherited by processes, highlights the dependency between processes which are responsible for a particular request.
We can look at this as a way of coloring a process dependency subgraph which is involved in responding to a particular request, and finding the process which creates the most unhappiness value at the current time.
The objective is to minimize system unhappiness.
The very first idea to achieve this is to find the request which creates the maximum aggregated unhappiness and allocate the CPU to the processes responsible for serving that request.
In theory we can achieve this by performing the following steps.

\begin{enumerate}
\item We assume that, there are two queues   in the system. 
 One for unhappy processes which is called $Q_{0}$ and the second queue for regular processes which is called $Q_{1}$.
 \item All processes with nonzero unhappiness values are placed in $Q_{0}$ and processes with zero unhappiness values are placed in $Q_{1}$.
 \item $Q_{0}$ has an absolute higher priority than $Q_{1}$, so there should be no processes in $Q_{0}$ before processes in $Q_{1}$ can be executed.
 \item In order to determine which process to execute next from $Q_{0}$, the scheduler first computes the unhappiness values for all pending requests in the system.
 \item Based on the requests' unhappiness values computed in the previous step, the scheduler chooses the request with the highest unhappiness value.
 \item The scheduler then, checks all processes responsible for that request. In other words it checks the dependency subgraph of the processes that are servicing that particular request and chooses the process with the highest unhappiness value to run on the CPU.
 \item Processes in $Q_{1}$ are executed based on a regular system scheduling algorithm for example Linux CFS scheduling algorithm.
 \item When a new process is created it should be given a nonzero unhappiness value so that it has a chance to start faster, then if it does not serve requests, it is moved to $Q_{1}$.

\end{enumerate} 

\section{Request Based Priority Elevation for CFS}
\label{sec-requestbased}
Unfortunately, implementing the simplest version of the proposed algorithm in Section \ref{sec-unhappy-shced-propos} requires that the scheduler detects all requests to all processes, and also requires detecting responses from processes to customers.
At present detecting a response from a process to a particular request seems to be impossible without process cooperation.
As a result of facing these difficulties, and in order to create a proof of concept implementation, we decide to take a simplified approach.
We begun by observing a typical Linux desktop which was also configured as a small web server.
This system had a Linux kernel version 2.6.31 and hence it uses CFS scheduler.
We used SystemTap \cite{SystemTap} and strace \cite{strace} to trace system calls and interprocess communications.
We observed that most of the requests from a desktop user is passed to processes through UNIX sockets.
Desktop components also mostly use UNIX sockets to communicate with each other \cite{dbus,DCOP}.
Requests from remote customers also enter the system through network sockets.
Based on these observations we propose a simple priority elevation technique in Linux kernels with CFS scheduler to approximately minimize system unhappiness as defined in Section \ref{sec-unhappyModel}.
\begin{enumerate}
 \item Assume that each process receives the incoming requests through a non-zero socket read.
 \item Since a pending request increases system unhappiness, whenever a process receives a request (non-zero socket read), scheduler should increase its priority or CPU share.
 \item The process should be able to maintain it's elevated priority until it sends a response to the customer. But, as we can't detect the exact  time when it sends a response back to the customer, scheduler decays the elevated priority in time.
 \item In order to minimize interaction with regular system scheduling, we change the amount of priority elevation and decaying speed based on the system load.
As the system load increases, an eligible process receives higher priority and retains this higher priority for a longer time.
This is based on the fact that when a Linux OS with CFS scheduler has a higher load, each process receives a smaller share of the CPU time \cite{CFS2-IMolnar,CFS-Mingo}.
So, in order for a given process that receives a request to be able to respod to the request as if there is little or no load on the system, it should receive a larger share of CPU time relative to other processes.
By increasing its priority more aggressively under heavier loads, we allocate more CPU time to such a process, as a result, it has a better chance to finish its computation and return a respond to the request in a shorter time interval.
\end{enumerate}
In the rest of this paper we refer to this method by its abbreviation CFS/RBPE.

\subsection{Unhappiness Values for CFS/RBPE}
\label{UnhapVal-CFS-RBPE}
In this subsection we compute theoretical unhappiness values for the priority elevation technique as we did for other schedulers in Section \ref{sec-comparison-theory}.
Again we consider two scenarios under this scheme and compute the amount of unhappiness observed by the customer.
The assumptions are mostly the same as what we assume in Section \ref{CFS-unhap-calc}.
There are $N-1$ tasks in the run queue of a single processor machine.
Task $P_{j}$ is an interactive job which is sleeping.
A customer sends a request $R$ to $P_{j}$. 
It wakes up and is inserted in the run queue.
Assume that priority elevation mechanism detects the request and uses negative nice value $-nice$, such that  $-15  \leq -nice < 0 $ to increase $P_{j}$'s priority.
Assume that elevated priority decays with a speed of 1 nice level per each $d$ seconds.
$P_{j}$ needs $\tau$ seconds to finish computation and return a response to the customer, and
$d < \tau$. As we explained in Section \ref{CFS-unhap-calc}, each negative nice level increases the task's weight to 1.25 times its previous value.
So if there are $N$ tasks including $P_{j}$ in the run queue and $P_{j}$ has a negative nice value $-nice$ then the following relations hold:
\begin{align}
\label{eq-RBPE-CPUshare-nice}
 &q(P_{j},N)=\frac{1.25^{nice}\hspace{10pt}sch\_lat}{N-1+1.25^{nice}}\nonumber\\
 &q(P_{i\neq j})=\frac{sch\_lat}{N-1+1.25^{nice}}
\end{align}

Assume $\tau \leqslant \frac{1.25^{nice}\hspace{10pt}sch\_lat}{N-1+1.25^{nice}}$ then as CFS enqueues an interactive task into the left of R-B tree, it is possible that $P_{j}$ finishes computation and returns a response to the customer within this time period. This means that, it is possible that the customer observes zero unhappiness.
\begin{equation}
\label{eq-RBPE-unhap-final-min}
 U^{R}_{min} = 0
\end{equation}

Now, let us consider a typical scenario where $\tau \geqslant \frac{1.25^{nice}\hspace{10pt}sch\_lat}{N-1+1.25^{nice}}=q$ and $P_{j}$ also needs a service from process $P_{s}$.
We also assume that $P_{j}$ first requires $\tau_{1} \approx Z_{1}q$ seconds before it blocks for service from $P_{s}$. $P_{s}$ requires $\tau_{2} \approx Z_{2}q$ to provide service to $P_{j}$, and eventually $P_{j}$ requires another $\tau_{3} \approx Z_{3}q$ to return a response to the customer. Also assume that $d \leqslant q$, so $P_{j}$ and $P_{s}$ nice levels increase almost every time it is rescheduled after it is set to $-nice$ by the scheduler. For simplicity we also assume that $Z_{1},Z_{2},Z_{3}\leqslant nice+1$. Based on these assumptions the amount of unhappiness observed by customer is:
\begin{align}
 U^{R} &= (\sum_{i=0}^{Z_{1}-2} \frac{(N-1)sch\_lat}{N-1+1.25^{(nice-i)}} - \sum_{i=1}^{Z_{1}-1} \frac{1.25^{(nice-i)}}{N-1+1.25^{(nice-i)}})\nonumber\\
&+(\sum_{i=0}^{Z_{2}-2} \frac{(N-1)sch\_lat}{N-1+1.25^{(nice-i)}} - \sum_{i=1}^{Z_{2}-1} \frac{1.25^{(nice-i)}}{N-1+1.25^{(nice-i)}})\nonumber\\
&+(\sum_{i=0}^{Z_{3}-2} \frac{(N-1)sch\_lat}{N-1+1.25^{(nice-i)}} - \sum_{i=1}^{Z_{3}-1} \frac{1.25^{(nice-i)}}{N-1+1.25^{(nice-i)}})
\end{align}

Assuming the total time needed to return a response to the customer is $\tau$, then we have $\tau = \tau_{1}+\tau_{2}+\tau_{3}$. Note that the CPU time that $P_{j}$ and $P_{s}$ receive is approximately the total execution time that they need to return a response to the customer request, so:
\begin{equation}
\label{eq-tau-in-RBPE-typical}
\tau=\sum_{i=1}^{Z_{1}-1} \frac{1.25^{(nice-i)}}{N-1+1.25^{(nice-i)}}+\sum_{i=1}^{Z_{2}-1} \frac{1.25^{(nice-i)}}{N-1+1.25^{(nice-i)}}+\sum_{i=1}^{Z_{3}-1} \frac{1.25^{(nice-i)}}{N-1+1.25^{(nice-i)}}
\end{equation}
And we can write $U^{R}$ as the following:

\begin{equation}
\label{eq-RBPE-unhap-typ-final}
 U^{R}=\sum_{i=0}^{Z_{1}-2} \frac{(N-1)sch\_lat}{N-1+1.25^{(nice-i)}}+\sum_{i=0}^{Z_{2}-2} \frac{(N-1)sch\_lat}{N-1+1.25^{(nice-i)}}+\sum_{i=0}^{Z_{3}-2} \frac{(N-1)sch\_lat}{N-1+1.25^{(nice-i)}}-\tau
\end{equation}

Please note that the above calculations are based on the assumption that all communications/requests between processes are performed using network or UNIX sockets.
Based on our observation this is a valid assumption for most Linux desktop applications/components.
Another fact is that most transactions need more than just one read operation.
For example a simple click on a link in a web browser causes many UNIX socket read system calls both in the web browser and the X server before a new page is displayed on the screen.
This in effect causes the active processes in the transaction (in this case web browser and X server) to receive the $-nice$ value multiple times.
It means that in practice, they have higher priority for a longer period of time than what we compute here.

\subsection{CFS vs CFS/RBPE}
\label{Discus-CFS-CFSprio}
In this subsection we compare and discuss unhappiness computations for CFS scheduler and CFS with RBPE.
As we see in sections \ref{CFS-unhap-calc} and \ref{UnhapVal-CFS-RBPE} the minimum unhappiness values for the best scenarios in both cases are zero.
So at the very minimum we can see that in theory our proposed RBPE scheme does not make the situation worse.
We can note that there are two conditions for the best case scenarios with the resulting zero unhappiness.
These conditions are:
\begin{align}
\label{eq-tau-nonequalities}
\tau &\leqslant \frac{sch\_lat}{N}  \hspace{20pt} For \hspace{5pt} CFS.\\
\tau &\leqslant \frac{1.25^{nice}\hspace{5pt}sch\_lat}{N-1+1.25^{nice}} \hspace{20pt} For \hspace{5pt} CFS/RBPE.
\end{align}

As $\frac{1.25^{nice}}{N-1+1.25^{nice}} \geqslant \frac{1}{N}$
clearly under CFS/RBPE, $P_{j}$ has more time to respond to the request before it is preempted than under CFS. So, there is higher possibility under CFS/RBPE that a request being responded without encountering any unhappiness.

For typical scenarios in both cases again we have:
\begin{align}
 U^{R} &= (N-1)(\frac{\tau}{sch\_lat} -3) -\tau  \hspace{20pt}For \hspace{5pt}CFS and \\
U^{R}&=(N-1)sch\_lat(\sum_{i=0}^{Z_{1}-2} \frac{1}{N-1+1.25^{(nice-i)}}+\sum_{i=0}^{Z_{2}-2} \frac{1}{N-1+1.25^{(nice-i)}}\nonumber\\
&+\sum_{i=0}^{Z_{3}-2} \frac{1}{N-1+1.25^{(nice-i)}})-\tau \hspace{20pt}For \hspace{5pt} CFS/RBPE
\end{align}
In the CFS/RBPE case, as a result of higher priority for $P_{j}$ and $P_{s}$, other tasks have less CPU time, so the resulting $U^{R}$ value is less than that of CFS case.

\section{Implementation}
\label{sec-implemetation}
While we were using SystemTap \cite{SystemTap} to observe process/OS interactions and behaviors, we found it extremely powerful to write simple scripts which can be used with different kernel versions without almost any modification.
So, in order to implement a simple Request Based Priority Elevation (RBPE) mechanism as a proof of concept for our customer appeasement, we decided to use SystemTap in its Guru mode.
When used in this mode, SystemTap enables parsing of expert-level constructs like embedded C.
So it basically enables us to write C code and insert it into the kernel as a kernel module, effectively modifying a running kernel without directly modifying kernel source code or recompiling it.


We use systemTap to create a list of process (PIDs) that recently have called a socket related receive/read system call with nonzero return value.
Assuming this call means that the associated process has received either a direct or indirect request from a customer, we increase its priority.
The exact negative nice value used to increase the process priority depends on the system load. 
The higher the system load the lower the negative nice value.
The applied negative nice value along with a time stamp is saved in a list with the associated process PID.
We call this list elevated priority process list (EPPL).
There is a time delay after that we increase the negative nice value of the processes which are in the EPPL, effectively reducing their priority.
The exact value of this time delay also depends on the system load.
During our process behavior observation period, we noticed that, the majority of processes waiting for an input, use the poll system call periodically.
We changed the poll system call and use it as a point to check the current system status and update the state of the processes which are in the elevated priority list. Each time a poll is called, we check EPPL and increase the nice value by one for each process that has passed its delay time. 
We then enter the new nice value with a new time stamp into the elevated process list.
For each process in the list, this action will continue until the nice value becomes zero, at that point the process is deleted from the list.
We adjust the values of initial nice values and delay time, based on the system load during poll system calls.
If system load is very low, the RBPE mechanism does not interfere with the regular CFS scheduling decisions, but, as the system load inceases it interferes aggressively, as mentioned earlier. 
Table \ref{tab-sysTapValues} shows the initial nice vales and delay times at different system loads in the current version (0.5) of our script.

\begin{table}[t]
\begin{center}
\label{tab-sysTapValues}
\begin{tabular}{| c | c | c | c | }
\hline
avenrun[0] $\leq$  & \emph{-nice1} (UNIX sockets) & \emph{-nice2} (Network sockets) & Time Delay (ms) \\
\hline
1600 &  0 & 0 & 0 \\ \hline
3000 & -1 & 0 & 200 \\ \hline
5000 & -2 & -1 & 300 \\ \hline
8000 & -4 & -2 & 400 \\ \hline
12000 & -6 & -3 & 500 \\ \hline
16000 & -7 & -4 & 600 \\ \hline
$>$ 16000   & -15 & -5 & 600 \\ \hline

\end{tabular}
\end{center}
\caption{RBPE script, nice and time delay values. avenrun[0] is a kernel variable which represents system's 1 minute load.}
\end{table}

In our implementation we discriminate local and remote users by giving higher priority to processes receiving requests through UNIX sockets relative to those processes receiving requests through network sockets.
This is implemented by using two initial negative nice values.
One with a lower value for processes that use local UNIX sockets and one with higher value for processes that use network sockets to receive requests.
All of these values are presented in Table \ref{tab-sysTapValues}.


Although this method of implementation might have a higher overhead, but as we see in Section \ref{sec-experiments}, it has very promising results for real world applications.

\section{Experiments}
\label{sec-experiments}
To test our CFS/RBPE technique, we performed multiple tests.
As we intend to show that our scheduling paradigm does not focus on interactive applications, we have performed server based performance tests as well as regular interactivity/multimedia tests.
Server based tests include Apache web server \cite{apache} performance test and Mysql data base server \cite{Mysql} performance tests.
We choose these two servers as they are very popular.
In fact many Linux based servers use Apache, Mysql and PHP to support different weblog, wiki or multimedia hosting services. 

\subsection{Hardware/Software Set up}
All tests are performed on an IBM (R) IntelliStation M Pro with Intel \emph{P4} 2.8GHz CPU and 1GB of RAM running Fedora 12 with Kernel 2.6.32.

In order to simulate different background system loads we compile Linux kernel and use different \emph{-j} values with the \emph{make} command to initiate different parallel compilations.

\subsection{Apache Web Server Response Test}
\label{sec-apache}
\begin{figure}[t]
\label{fig-apache1}
\centering \includegraphics[width=140mm,height=100mm]{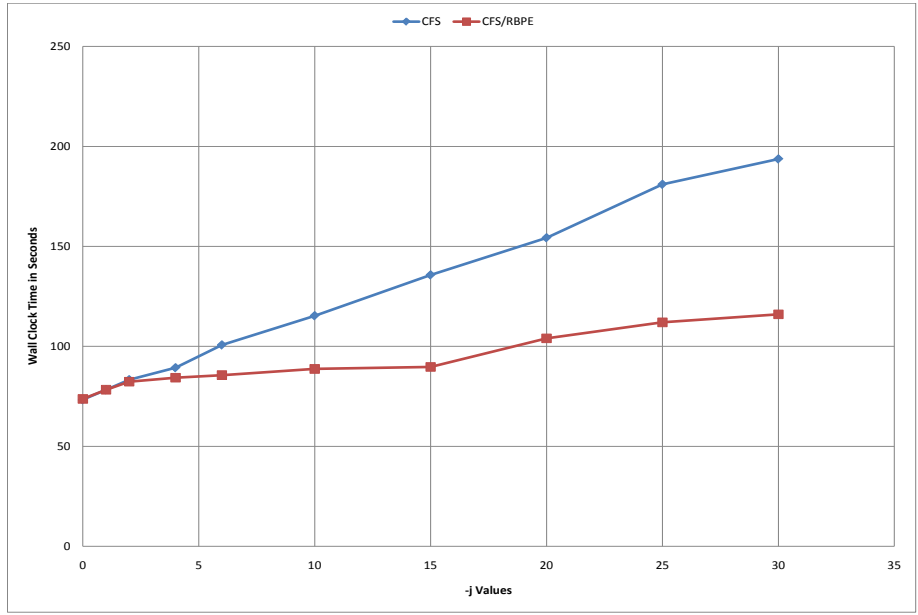}
\caption{Download times in Seconds from an Apache web server under different simulated load values.}
\end{figure}
As a first test to measure server based application performance we measure the response time of Apache web server under different system loads.
The web server hosts a directory structure of files.
We use a second machine and \emph{wget} command to download the complete directory structure from the web server.
We use shell \emph{time} command to measure wall clock download times under different simulated load conditions.
For each simulated load value we repeat the experiment 3 times and compute the average download time for that system load.
As we see in Figure \ref{fig-apache1}, both CFS and CFS/RBPE have the same response time  for \emph{-j} values up to 2 after that point CFS/RBPE has consistently lower response times.
When \emph{-j} value equal to 30 is used Apache web server is almost 1.5 times slower when it is run under CFS than under CFS/RBPE.

\subsection{SysBench Mysql Data Base Test}
\label{sec-Mysql}
\begin{figure}[t]
\label{fig-Sysbech-mysql}
\centering \includegraphics[width=140mm,height=110mm]{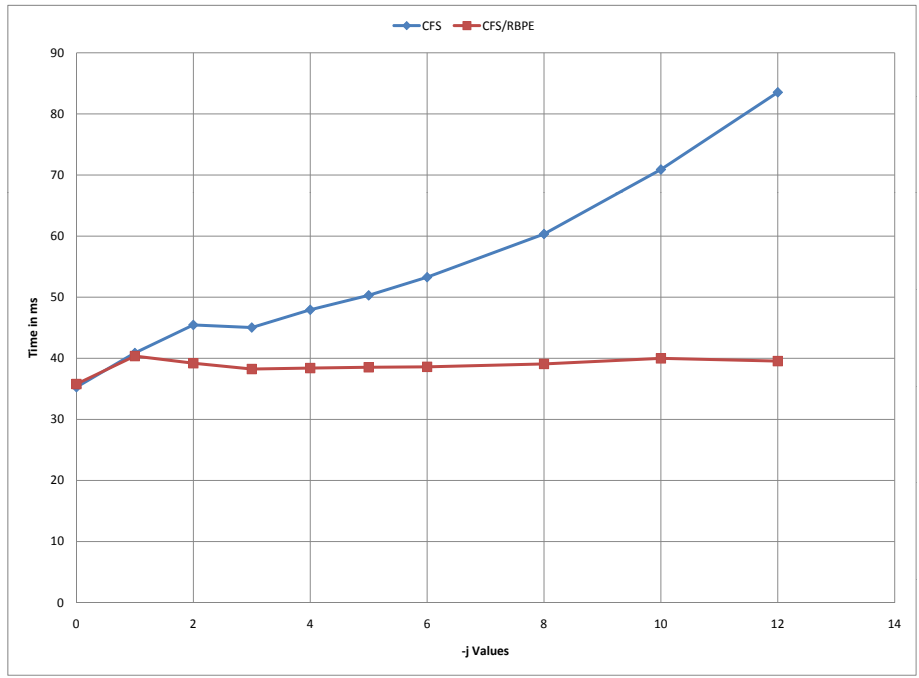}
\caption{It shows the average transaction times in milliseconds under different simulated load conditions. The X values are \emph{-j} values passed to the make command to specify number of parallel compilations to initiate.}
\end{figure}

As a second server based software, we test Mysql response times under different background load conditions. 
We use Stench \cite{SysBench} to evaluate and compare Mysql data base performance under Linux CFS \cite{CFS2-IMolnar,CFS-Mingo} and CFS/RBPE scheduling policy.
Stench is a multi-threaded benchmark tool which can evaluate OS parameters important for a system running on a data base server under heavy load.
We use Stench default parameters for its OLTP complex configuration.
During each experiment Stench executes many read and write transactions on the Mysql data base server, and then gives the average transaction time.
Each experiment runs for 120 to 250 seconds under different simulated load conditions.
We change experiment times in order to have around 3000 transactions per each experiment.
The reason is that, when system load increases the total number of transactions in a fixed time interval decreases.
So we increase the experiment time to have almost equal number of transactions for each experiment.
This gives us a better average transaction time in all experiments.
Figure \ref{fig-Sysbech-mysql} shows the average transaction time in milliseconds under different simulated load conditions.
Load conditions were simulated by running a Linux kernel compilation job with different \emph{-j} options to the \emph{make} command as in the previous test in Section \ref{sec-apache}. 

As we see in Figure \ref{fig-Sysbech-mysql}, the average Mysql tarnsaction time under CFS/RBPE increases first as \emph{-j} value increases to 1 (meaning from no load to one parallel kernel compilation) but, then it decreases when parallel compilation increased to two and three. 
This is because RBPE does not interfere with the usual CFS scheduling when system load is low. 
When the system load increases RBPE is involved and as we see in Figure \ref{fig-Sysbech-mysql}, it boosts Mysql performance so that its average transaction time is almost constant near 40 milliseconds.
In contrast, under CFS scheduling, Mysql average transaction time increases as system load increases. 
When we use a \emph{-j} value of 12, the average transaction time reaches 85 milliseconds which is more than two times that of CFS/RBPE.

The results of the experiments in this section and the previous section  indicate that the proposed scheduling paradigm in this paper is not a method to just boost desktop interactive applications response time.
This mechanism can boost the performance of any request/response based transaction in the system.

\subsection{Interactivity/Multimedia Test}
\label{sec-mplayer}
In order to test and compare the performance of interactive/multimedia applications under CFS and CFS/RBPE schedulers, we use \emph{mplayer} in benchmark mode.
In this mode \emph{mplayer} prints out the number of dropped frames and average frame rate after it finishes playing a multimedia file. 
We use a short mpeg clip of size 352 x 288 pixels, which runs for about 150 seconds.
The clip frame rate is 25 frames per second.
Again we simulate the system load with parallel compilation of Linux kernel and use \emph{make -j} with different values for \emph{-j} to control the number of parallel \emph{makes}.

\begin{figure}[t]
\label{fig-mplayer-drop}
\centering \includegraphics[width=140mm,height=110mm]{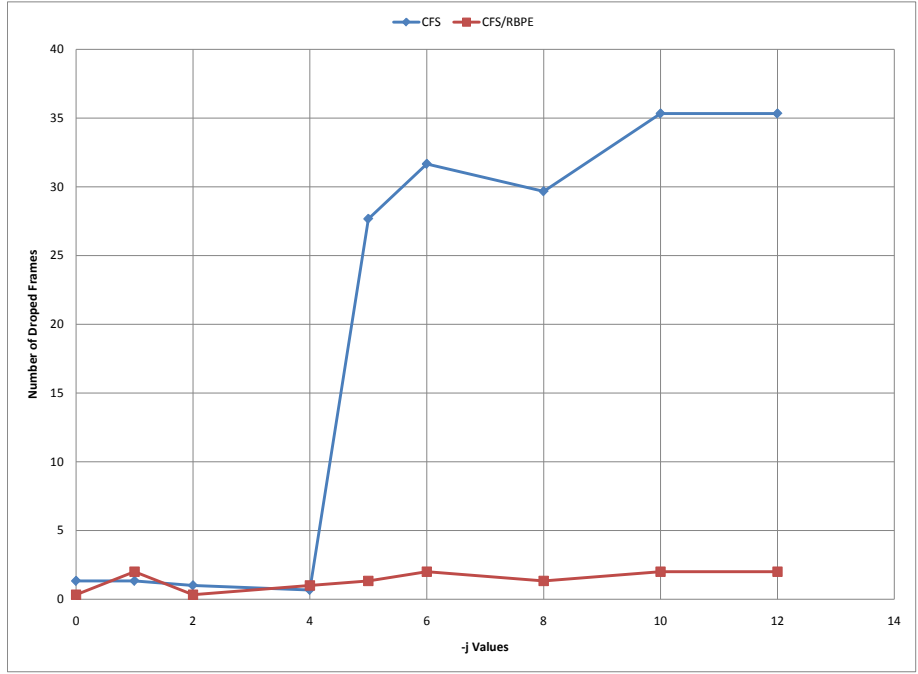}
\caption{This figure compares the frame rate drop when mplayer is playing an mpeg movie clip and a simulated background load was increasing.}
\end{figure}

For each \emph{-j} value the experiment is repeated three times and the average value of dropped frames is depicted in Figure \ref{fig-mplayer-drop}.
As we see in this graph under CFS/RBPE the frame drop is almost zero for all load values up to \emph{-j 12}.
Under CFS the number of frame drops significantly increases after \emph{-j 4}.

\begin{figure}[t]
\label{fig-mplayer-frameRate}
\centering \includegraphics[width=140mm,height=110mm]{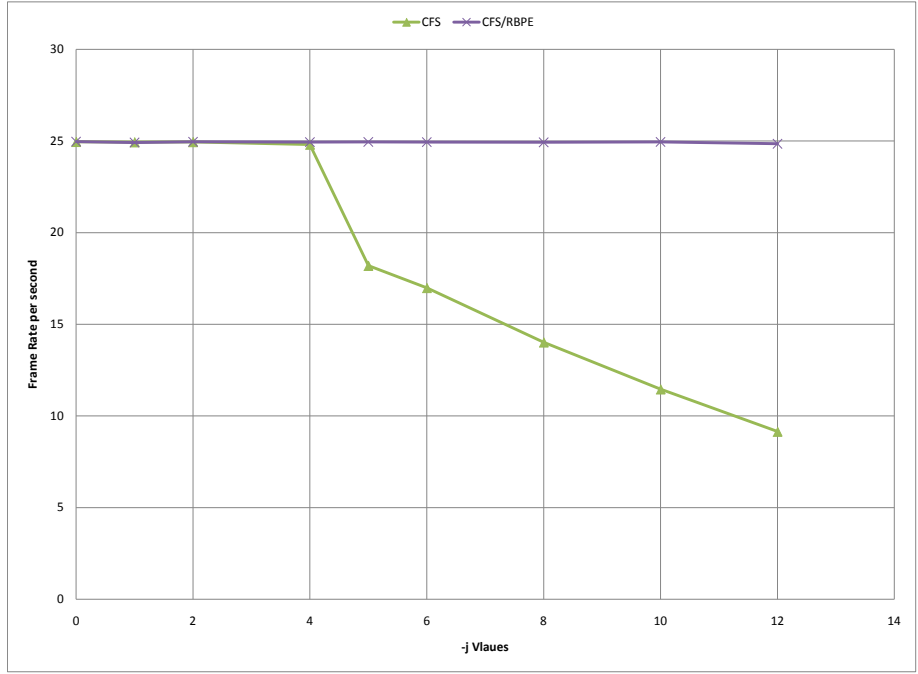}
\caption{Demonstrates frame rate change due to system load under CFS scheduler and CFS/RBPE.}
\end{figure}

We also depict the average frame rate of \emph{mplayer} for CFS and CFS/RBPE under different simulated load levels.
As we see in Figure \ref{fig-mplayer-frameRate}, \emph{mplayer} frame rate drops to almost 8 frame per second under CFS when 12 parallel compilation is running, while at the same load level CFS with RBPE shows almost no frame rate reduction.

This experiment shows the effectiveness of CFS/RBPE on a typical multimedia or streaming application.
As Figures \ref{fig-mplayer-drop} and \ref{fig-mplayer-frameRate} indicate, basically the movie is not viewable when \emph{-j} value reaches 5 on our system with CFS scheduling.
In contrast a viewer can still enjoy watching a movie on the same system if CFS/RBPE scheduling is used even if \emph{make -j} with value of 12 is used for compiling a linux kernel at the same time.

\section{Concluding Remarks}
\label{sec-conclusion}
In this work we introduce a new policy for CPU scheduling.
This policy is based on tracking requests sent by customers to different processes and their response to the requests.
We assume that a computer system should allocate its resources such that the customers do not experience excessive delays. 
We have defined a model which can be used to analyze and compare different scheduling algorithms based on this assumption.
This model considers delays resulted from processes dependencies. 
When a requests arrives at the system, one or more processes are responsible to execute the request and return a response to the customer.
We detect the request and the processes which are responding to the request by tracking interprocess communications.
We have a minimal implementation on top of Linux CFS scheduler \cite{CFS2-IMolnar,CFS-Mingo} as a proof of concept which increases the priority of the processes involved in a response to a customer request.
Our experiments show that this mechanism is not only effective for improving interactive/desktop applications performance under heavy system load, it is also effective for improving server applications under heavy background load.
Experiments with Apache web server and Mysql data base server in Sections \ref{sec-apache} and \ref{sec-Mysql}, show significant performance boost for these server applications under heavy background load.
A server background load may be the result of disk indexing, data base indexing, log rotation, log analysis, etc.

One of our goals is to make the implementation simple, easily portable to different Linux kernels and distributions, and easy to use for a novice user.
In order to achieve this, we use SystemTap \cite{SystemTap}.
SystemTap has made our implementation a simple script which can be run on any SystemTap equipped distribution with a compatible kernel without the need to recompile and install a new kernel.
There has been a debate and disagreement among kernel development community on whether to support SystemTap or not \cite{SystemTapDebate}.
If the support for SystemTap is dropped by its main developers then as far as we know there is no alternative for a fast and easy to use implementation as we have done in this work.

Some of the possible future works are:
\begin{enumerate}
 \item Integration of the implementation with the Linux kernel instead of using SystemTap.
 \item Extending the proposed CPU scheduling paradigm to disk scheduling in the sense that higher priority process also get higher priority disk access.
\item Studing the effects of adding other interprocess communications like pipes to the implementation.
\item Adding other mechanisms to detect arrival of a request to the system.
\item Finding ways to give different weights to different requests from a specific customer.
\item Studing the usage of proposed mechanism in managing resources used by different virtual machines on one real machine. 
\end{enumerate}

\bibliographystyle{plain}
\bibliography{reference}
\end{document}